\documentclass[a4paper,11pt]{article}
\pdfoutput=1 

\usepackage{jheppub} 

\usepackage[T1]{fontenc} 

\usepackage{graphicx}
\usepackage{color}
\usepackage{bm}
\usepackage{amsmath}
\usepackage{amsfonts}
\usepackage{amssymb}
\usepackage{mathrsfs}
\usepackage{caption}
\usepackage{subcaption}
\usepackage{enumerate}

\def\be{\begin{equation}}
\def\ee{\end{equation}}
\def\eq#1{{Eq.~(\ref{#1})}}

\def\sect#1{{Section~\ref{#1}}}
\def\fig#1{{Fig.~\ref{#1}}}

\title{It's a dark, dark world: Background evolution of interacting $\phi$CDM models beyond simple exponential potentials}

\author{Suprit Singh}
\author{and Parminder Singh}

\affiliation{Department of Physics and Astrophysics,\\University of Delhi, New Delhi 110 007, India}

\emailAdd{ssingh2@physics.du.ac.in}
\emailAdd{psingh@physics.du.ac.in}

\abstract{We study the background cosmological dynamics with a three component source content: a radiation fluid, a barotropic fluid to mimic the matter sector and a single scalar field which can act as dark energy giving rise to the late-time accelerated phase. Using the well-known dimensionless variables, we cast the dynamical equations into an autonomous system of ordinary differential equations (ASODE), which are studied by computing the fixed points and the conditions for their stability.  The matter fluid and the scalar field are taken to be uncoupled at first and later, we consider a coupling between the two of the form $Q = \sqrt{2/3}\kappa\beta\rho_m\dot{\phi}$ where $\rho_m$ is the barotropic fluid density. The key point of our analysis is that for the closure of ASODE, we only demand that the jerk, $\Gamma = V V''/V'^2$ is a function of acceleration, $z = - M_p V'/ V$, that is, $\Gamma = 1+ f(z)$. In this way, we are able to accommodate a large class of potentials that goes beyond the simple exponential potentials. The analysis is completely generic and \emph{independent} of the form of the potential for the scalar field. As an illustration and confirmation of the analysis, we consider $f(z)$ of the forms $\mu/z^2$, $\mu/z$, $(\mu-z)/z^2$ and $(\mu-z)$ to numerically compute the evolution of cosmological parameters with and without coupling. Implications of the approach and the results are discussed.}

\begin{document}
\maketitle
\flushbottom

\section{Introduction}

The evidence of dark energy being present in the cosmological soup comes from various observations such as the distance measurements from type Ia supernovae \cite{sn1}, baryon acoustic oscillations \cite{bao}, measures of the Hubble constant \cite{hubble} and the precision measurements of cosmic microwave background \cite{cmb}. The data from these observations consistently favours the spatially flat $\Lambda$CDM model with a currently accelerating phase. Here, the cosmological constant $\Lambda$ is used as an ingredient responsible for dark energy with the equation of state parameter $w_\Lambda = -1$ to produce the late-time accelerating phase. However, an exceptionally small value of the cosmological constant with $\Lambda L_p^2 \approx 10^{-122}$ (fine-tuning) and the relatively equal order of magnitude values of $\Omega_\Lambda \approx 0.7$ and $\Omega_m \approx 0.3$ (coincidence) suggest that dark energy could be more than just having a cosmological constant in general relativity\footnote{It should be noted that apart from these concerns, there are no practical issues concerning the gravitational effects and cosmological implications of having a cosmological constant. See ref.~\cite{padilla} for a field theoretic viewpoint of the cosmological constant problem.}.

There are many possibilities that have been explored to account for the late-time acceleration of the universe (for a non-exhaustive sample see refs.~\cite{sahni1, paddy, bludman, ed, sahni2,  frieman, caldwell, tsujikawa, amendolabook, wang, clifton, bamba}). A popular and simplest approach is using a light scalar field termed ``quintessence'' as a candidate for dark energy.  In the quintessence models, the scalar field is either taken to be slowly rolling along a shallow potential (``thawing'' class) \cite{thawing} or else with a ``scaling'' behaviour \cite{scaling1,scaling2,scaling3,scaling4} (which contains ``tracker'' as a special case) in which it tracks the dominant fluid component and in the later stage exits the scaling regime with a quasi-constant equation of state parameter. The form of the self-interacting potentials for the scalar field is usually phenomenological or else is motivated as arising from the theories with extra degrees of freedom such as Kaluza-Klein reduction of a higher dimensional theory \cite{highdim1,highdim2} or from string theoretic \cite{string} considerations. Further we can also consider, in principle, an interaction between dark energy (the scalar field here) and matter fluid (including dark matter). Such an interaction is again either introduced phenomenologically to address the coincidence problem (also affecting structure formation in new ways) or can be motivated by considering modified gravity\footnote{In some circles, the presence of dark energy is synonymous with modifying the gravity sector rather than having an extra source in Einstein gravity. However, conformal transformations can bring most of the modified gravity actions back to the canonical Einstein form with extra source degrees of freedom. It is then a matter of debate to ask which one of these conformally related frames is more physical or fundamental.} models in the Einstein frame \cite{cup1,cup2,cup3,cup4,cup5,cup6}. The couplings that arise from such considerations affect the matter sector with an effective metric of the form \cite{disformal1,disformal2,disformal3,disformal4}:
\be
\bar{g}_{ab} = C(\phi) g_{ab} + D(\phi) \phi_a\phi_b
\ee
where $C(\phi)$ and $D(\phi)$ refer to the conformal and the disformal part of the coupling respectively. With these model inputs, we can study the background and perturbed cosmological dynamics for parameter estimation as well as to infer the concordance with observations. The background evolution is usually and effectively studied by transforming the cosmological dynamical equations to an equivalent non-linear autonomous set of differential equations \cite{intro}. The only tractable analytical approach is to look at the fixed points of the equivalent system and their stability conditions. These fixed points can be associated with different cosmological phases which can be either early-time (source), transients or late-time (sink) solutions. The numerical integration of the differential equations with given initial conditions reveals the phase space dynamics and the evolution of cosmological parameters. We can compute the background quantities such as the evolution of Hubble parameter, equation of state parameter, luminosity distances etc. and compare these with observations. The perturbed evolution requires running modified versions of popular CMB Boltzmann codes.

However, many subtleties hide in the modelling of the dark sector and one requires a careful analysis of each step one at a time which brings us to the motivation (for yet another paper) and the specific point of view adopted here. In this work,
\begin{itemize}

\item We study the background cosmological evolution in a fairly generic setting. The analysis encompasses a large class of potentials and the interacting dark sector models including those arising out of the modification of gravity sector all (except disformal) \emph{in one go}.

\item We consider a three component system with radiation, a barotropic fluid and the quintessence scalar field with a (quite) generic potential. This is as close as we can get to the actual composition of our universe that can be handled analytically.

\item We present a classification of a large number of potentials for the quintessence field beyond the simple exponential potential. There exist three main classes for all such potentials based on the properties of the function $\Gamma -1 = V V''/V'^2 -1$.

\item We then proceed to understand the cosmological dynamics through dynamical system analysis in terms of computing the fixed points, understanding their properties and stability conditions.

\item Finally, we resort to numerical techniques and show the evolution of cosmological parameters for a few examples in which tracking or thawing behaviour is observed.
\end{itemize}
Most of the previous studies on similar lines \cite{exact1,exact2} consider only two components -- the scalar field and a matter fluid -- focusing only on the late-time accelerated regime where the contribution of radiation is negligible. However even though it is subdominant at present, in order to obtain a correct tracking regime \cite{barrow} from early epochs, we have to include in radiation from the start. We have not seen a complete and generic analysis of three component system anywhere in the literature and it is therefore interesting to see what we can infer from it.

The outline of the paper is as follows: We introduce the machinery of an equivalent dynamical system in \sect{asode}, classification of the potentials in \sect{classpot} and briefly review the linear stability analysis in \sect{stability}. In \sect{fpa}, we compute the fixed points, the conditions of their existence and stability and cosmological implications with and without coupling. Using numerical techniques in \sect{numerics}, we consider some examples to demonstrate and confirm the analytical results obtained in the previous section before summarizing the conclusions in \sect{conc}.  

\section{Theoretical Setup}
\label{sec:theoretical}

In this section, we introduce the theoretical framework of the (interacting) quintessence models with a brief review of the dynamical systems analysis relevant for our purposes (see ref. \cite{intro} for an excellent introduction). We, first, develop the machinery to go from the cosmological dynamical equations to an equivalent system of dimensionless phase space variables. These satisfy an autonomous set of differential equations (ASODE). We then mention the rules involved in the stability analysis of a non-linear autonomous system.

\subsection{From cosmological dynamical equations to ASODE}
\label{asode}

We assume a spatially flat, homogenous and isotropic background which gives the diagonal Friedmann metric, $g_{ab} = (-1, a^2,a^2,a^2)$. Assuming the framework of general relativity and the sources,
\begin{itemize}
\item a scalar field $\phi$, minimally coupled to gravity: $\mathcal{L}_\phi = \dot{\phi}^2/2 - V(\phi)$ with the equation of state, $P_\phi = w_\phi \rho_\phi=(\gamma_\phi -1) \rho_\phi$ where $\rho_\phi = \dot{\phi}^2/2 + V(\phi)$ and $P_\phi = \dot{\phi}^2/2 - V(\phi)$,
\item a barotropic fluid (matter sector) with $p_m =  w_m\rho_m = (\gamma - 1) \rho_m$ and $1\leq \gamma<2$,
\item a radiation fluid for which $P_r = \rho_r/3$,
\end{itemize}
the dynamical equation for this metric is given by,
\be
\dot{H} = -\frac{1}{2M_P^2}\left(\dot{\phi}^2+\gamma\rho_m + \frac{4}{3}\rho_r\right)
\ee
where we have the standard definitions, $H \equiv \dot{a}/a$ and $M_P^2 \equiv 1/(8\pi G)$. With this, the background evolution is also required to satisfy,
\be
\label{frconstraint}
H^2 = \frac{1}{3M_P^2}(\rho_\phi +\rho_m + \rho_r) \hspace{5pt}\mathrm{or} \hspace{5pt} \Omega_\phi+\Omega_m+\Omega_r = 1; \hspace{5pt}\Omega_i = \frac{\rho_i}{3 H^2M_P^2}
\ee
as the flatness constraint. Further, the stress-energy tensor for the sources has to be divergence-less. In the absence of any coupling between the components, the stress-energy tensors for the components are divergence free individually. However, in the presence of any coupling, the divergence-free condition holds in an added manner. This leads to the following set of (continuity) equations:
\begin{align}
&\dot{\rho}_\phi + 3 H \gamma_\phi \rho_\phi= Q\label{phicty}\\
&\dot{\rho}_m + 3 H \gamma\rho_m = - Q\label{mattercty}\\
&\dot{\rho}_r + 4 H \rho_r = 0\label{radtncty},
\end{align}
where $Q$ denotes the coupling between the scalar field and the matter sector. One can construct the coupling purely on the basis that at least $Q$ should depend on the energy densities and other covariant quantities. As remarked before, we can also get a coupling between the scalar field and the matter sector by writing a modified (scalar-tensor) gravity theory in canonical Einstein form (see refs.~\cite{cup1,cup2}). Locally, the modified continuity equations imply the rate of energy exchange between the two components which can be classified as
\be
Q
\left\{
\begin{array}{c}
 > 0  \quad\quad\mathrm{dark~matter} \rightarrow \mathrm{dark~energy}\\
< 0 \quad\quad\mathrm{dark~energy} \rightarrow \mathrm{dark~matter}
\end{array}
\right.
\ee
The dynamical equation for the scalar field with this coupling is given by,
\be
\ddot{\phi} + 3 H\dot{\phi} + V'(\phi) = Q/\dot{\phi}
\ee
obtained by substituting for $\rho_\phi$ in \eq{phicty} and hence is not independent of the continuity equation. This completes the set of dynamical equations for our system. For our purpose of the tracking the behaviour of $\Omega_\phi$ and $w_\phi$, it is convenient to introduce the dimensionless, phase space variables \cite{variables1,variables2}:
\be
x^2 \equiv \frac{\dot{\phi}^2}{6 H^2M_P^2}; \hspace{5pt}y^2 \equiv \frac{V}{3 H^2M_P^2};\hspace{5pt}u^2 \equiv \Omega_r = \frac{\rho_r}{3 H^2M_P^2};
\ee
such that the dimensionless energy density of the field $\phi$ is given by $\Omega_\phi  = x^2 + y^2$ and $w_\phi = ( x^2-y^2)/( x^2+y^2)$. The Friedmann constraint can be written as
\be
\Omega_m = 1- x^2 - y^2 - u^2.
\ee
This constrains the physical phase space with the conditions,
\be
0 \leq \Omega_m \leq 1 \Rightarrow 0\leq x^2 + y^2 + u^2\leq1.
\ee
Besides, since $0\leq\Omega_\phi\leq1$, we have $0\leq x^2+y^2\leq1$ as well as $|x|\leq 1$. Additionally, with positive cosmic expansion, that is, $H\geq0$, we have $y\geq0$. Also, $0\leq\Omega_u\leq1$ gives $|u|\leq1$ but we shall see that the system of equations will be symmetric under $u\rightarrow -u$ and hence the physical phase space is defined by,
\be
\Psi_\phi := \left\{(x,y,u): 0\leq  x^2 + y^2 + u^2\leq1, 0\leq  x^2+y^2\leq1, |x|\leq 1, y\geq 0, u\geq0\right\}
\ee
which is a bounded upper semi-disk. This is enough for the potentials that are constant or belong to simple exponential class, $V (\phi) = a e^{\pm b\phi}$. For the potentials beyond these simple classes, we need to introduce more variables. The first such variable \cite{variableZ1,variableZ2} is,
\be
z \equiv -M_P\frac{V'}{V}
\ee
which vanishes for a constant potential and is constant for an exponential potential. In terms of these set of variables $\{x,y,u,z\}$, we can recast the dynamical equations as:
\begin{align}
\frac{dx}{d N}&=-3x+\sqrt{\frac{3}{2}} y^2 z+\frac{3}{2} x \left[(2-\gamma) x^2 + \gamma (1-y^2)+(4-3\gamma)\frac{u^2}{3}\right]+\frac{Q x}{\epsilon\dot{\phi}^2 H}\label{x}\\
\frac{dy}{d N}&=-\sqrt{\frac{3}{2}} x y z+\frac{3}{2} y \left[(2-\gamma) x^2 + \gamma (1-y^2)+(4-3\gamma)\frac{u^2}{3}\right]\label{y}\\
\frac{du}{dN}&=-2 u+\frac{3}{2} u \left[(2-\gamma) x^2 + \gamma (1-y^2)+(4-3\gamma)\frac{u^2}{3}\right]\label{u}\\
\frac{dz}{dN}&=- \sqrt{6}\, x z^2 (\Gamma - 1); \hspace{5pt}\Gamma = \frac{V V''}{V'^2}\label{z}
\end{align}
where we have now introduced $N =\int da/a$ as the time parameter which is also a monotonically increasing function. For the closure of the above set of equations, we now  demand that $\Gamma - 1 = f(z)$ where $f(z)$ is some arbitrary function of $z$. Otherwise, one needs to consider an additional ODE: $\Gamma_{,N} = \ldots$ and the phase space becomes more complicated. However, even within this constraint of the jerk, $\Gamma$ being some function of the acceleration $z$, we are able to accommodate a very large class of potentials beyond a simple exponential class. With this framework, we now wish to understand the dynamics of the variables, $\{x,y,u,z\}$ as per the above autonomous set of differential equations\footnote{The system,  $x_i' = f_i(x_j)$ forms an autonomous set of differential equations (ASODE) when the functions $f_i(x_j)$ do not depend on the evolution parameter explicitly.}(ASODE).

\subsection{Classification of potentials}
\label{classpot}

With the identification of $(\Gamma-1)$ with $f(z)$ in \eq{z}, we still have a large class of potentials at hand for the scalar field. The mapping of the potentials $V(\phi)$ with the function $f(z)$ and the form of the right hand side of \eq{z} i.e., $dz/dN = -\sqrt{6}xz^2f(z)$ provides a way to classify them as follows:

\begin{itemize}
\item Potentials for which $f(z_* = 0) = 0$, e.g.,  $f(z) = \mu z^\alpha$ or $(\sinh(\mu z))^\alpha$ with $\alpha\geq0$,
\item Potential for which $f(z_*) = 0$ with $z_* \neq 0$. In this case, we can have two further classifications.
\begin{itemize}
\item $z^2f(z)|_{z=0}\neq0$ e.g., $f(z) = (\mu -z^\alpha)/z^2$. Then $x=0$ is necessary for an allowed fixed point.
\item $z^2f(z)|_{z=0}=0$ e.g., $f(z) = (\mu-z^\alpha)$ or $ (\mu-z^\alpha)/z$.
\end{itemize}
\item Potentials for which $z_*$ is not finite e.g., $f(z) = \mu/z^\alpha$ with $0< \alpha\leq2$.
\end{itemize}
We can get the potentials $V(\phi)$ connected with an $f(z)$ easily by computing $V(z)$ and then substituting for $z(\phi)$. From the definition $z\equiv-M_P V'/V$ it is straightforward to see that,
\be
\frac{dz}{dV} = \frac{z f(z)}{V},
\ee
which gives
\be
\label{vz}
V(z) = V_0 \exp\left(\int\frac{dz}{zf(z)}\right)
\ee
and similarly we have
\be
\frac{dz}{d\phi} = -\frac{1}{M_p} z^2 f(z).
\ee
which solves to
\be
\phi(z) =  -M_p\int \frac{dz}{z^2f(z)}.
\ee
In principle, we can now invert the above expression to get $z(\phi)$ which can be substituted in \eq{vz} giving $V(\phi)$. However, it is important to note that the mapping between the potential and $f(z)$ is not completely unique. For example, in the case of $V(\phi) = V_0 e^{\alpha e^{-\beta\phi/M_p}}$ \cite{variableZ2}, we have $f(z) = \beta/z$ which contains no information about the parameter $\alpha$ in the potential. Thus, $\alpha$ does not affect the dynamics even though the potential has a very different behaviour for $\alpha<0$ and $\alpha>0$ with a given $\beta$. The dynamics is only sensitive to $f(z)$ and any parameter estimation run using the background level observational aspects will not determine the potential completely. Hence, we shall just focus on $f(z)$ in our analysis which in its simplest form can be written as $f(z) = \mu(a + b z^c)$ where the parameters $\{\mu, a, b, c\}$ are real.

\subsection{Linear stability analysis of an autonomous system}
\label{stability}

We can assess the dynamics of the system in the phase space by looking at the fixed or critical points for the ASODE and analysing them for their stability at the linear level. Essentially, this means that for a set of equations, $x_i' = f_i(x_j)$, we find the set of points, $\{(x^{(c)}_j)\}$ for which all $f_i(x^{(c)}_j)$ vanish simultaneously. The next step is to consider a small perturbation in the neighbourhood of each fixed point, that is, taking $x_i^{(c)}+\delta x_i$  and expanding $f_i(x^{(c)}_j+\delta x_j)$ to linear order thus linearizing our equations in order to judge the stability of the given equilibrium point. Then, depending on whether the linear perturbations decay(grow) with time $N$ or decay in one direction while grow in the other, the critical point can be a future (past) attractor or a saddle point. This inference essentially follows from the study of the eigenvalues $\lambda_i$ of the Jacobian,
\be
J = \left(\frac{\partial f_i}{\partial x_j}\right).
\ee

\begin{itemize}
\item If $\mathrm{Re}(\lambda_i)\neq0$ $\forall i$, the point is termed as a hyperbolic critical point.
\begin{itemize}
\item With $\lambda_i \in \mathrm{Reals}$ and
\begin{itemize}
\item[(a)] all $\lambda_i < 0$, the critical point is a stable node (future attractor) or a sink.
\item[(b)] all $\lambda_i > 0$, the critical point is an unstable node (past attractor) or a source.
\item[(c)] $\lambda_i < 0 <\lambda_j$ for some $i,j$ the critical point is a saddle point (marginally stable), that is, a source in one direction and sink in the other.
\end{itemize}
\item If eigenvalues are complex numbers with $\lambda,\lambda^* = \nu\pm i\omega$ and $\omega\neq0$, then if
\begin{itemize}
\item[(a)] $\nu< 0$, the critical point is a stable spiral with damped oscillations.
\item[(b)] $\nu > 0$, the critical point is an unstable spiral with growing oscillations.
\end{itemize}
\end{itemize}
\item If $\mathrm{Re}(\lambda_i) = 0$ for some $i$, the point is a non-hyperbolic critical point. In this case if the imaginary part is finite, it is associated with free oscillations. If such a point has at least one positive eigenvalue, it is an unstable point. Otherwise, one has to go beyond the linear stability theory and use the centre manifold theory in order to infer the stability criterion.
\end{itemize}
By knowing the critical points and their nature in the phase space corresponding to a given cosmological model, we can correlate the attractors with cosmological solutions that can decide the end point and/or the origin of the cosmic evolution. The meta-stable saddle equilibrium points which are not local extrema can be associated with equally important transient cosmological solutions. Thus, in a convenient trade-off, instead of studying the cosmological dynamics given by the Hubble parameter, $H(t)$ and the densities of the sources $\rho_i(t)$, we opt to study the equilibrium points of an equivalent ASODE in the phase space of the cosmological model. 

\section{Fixed point analysis and cosmological implications}
\label{fpa}

We now turn to the computation of fixed points, their existence and the stability conditions. We first look at the case without any coupling between matter and the scalar field after which we look the effects with the coupling turned on. We also give the cosmological implications of each of the fixed points which can be associated with source/sink/transient cosmological behaviour according to their stability.

\subsection{Dynamics without any interaction between the components}

Consider the case when there is no interaction between matter and the scalar field, that is, taking $Q = 0$ in \eq{x}.
It is clear from \eq{z} that for $z$ to be stationary, we need either of $x$, $z$ or $f(z)$ to vanish. Hence we can look at the fixed points case by case, that is, when (i) $z=0$, (ii) $z=z_*$ for which $f(z_*) = 0$ and (iii) $z=z_a$ (arbitrary) which requires $x=0$. We find that there are 12 fixed points in this system. These along with their eigenvalues in the linearised system are tabulated in Table~\ref{table1}. The conditions for their existence, nature (stability) and the cosmological parameters $w_{\mathrm{eff}}$, $\Omega_\phi$, $w_\phi$ and $\Omega_m$ at those points are given in Table~\ref{table2}.

\begin{table}[t!]
\centering
\begin{tabular}{|c|c|c|c|c|c|c|c|c|c|c|c|}
\hline\hline
Points        &$x$ &$y$ &$u$ &$z$                                                          &$\lambda_1$ &$\lambda_2$&$\lambda_3$&$\lambda_4$                          \\
\hline\hline
P$_1$         & 0  &1  &0  &0                                &-2 &$-3\gamma$ &\multicolumn{2}{c|}{$-\frac{3}{2}\pm\frac{3}{2}\sqrt{1-\frac{4}{3}z^2f(z)}|_{z=0}$}            \\ \hline
P$_2^\pm$     &$\pm1$  &0 & 0 &0                                                                                  &3 &1 &$\kappa_1$ &$3a$                                                               \\ \hline
P$_3$         & 0 &0 &0 &$z_a$                                                                                    &0 &$\frac{3\gamma}{2}$ &$-\frac{3a}{2}$ &$-\frac{b}{2}$                   \\ \hline
P$_4$         & 0  &0 &1 &$z_a$                                                                                       &0 &2&-1&$b$                                                            \\ \hline
P$_5$         &0 &0 &0 &$z_*$                                                                                     &0 &$\frac{3\gamma}{2}$ &$-\frac{3a}{2}$ &$-\frac{b}{2}$                   \\ \hline
P$_6$         &0 &0 &1 &$z_*$                                                                                     &0 &2 &-1 &$b$                                                             \\ \hline
P$_7^\pm$     &$\pm1$ &0 &0 &$z_*$                                                                                &1 &$3a$ &$\frac{6\mp\sqrt{6}z_*}{2}$ &$\mp\sqrt{6}z_*^2df_*$              \\ \hline
P$_8$         &$\frac{z_*}{\sqrt{6}}$ &$\sqrt{1-\frac{z_*^2}{6}}$ &0 &$z_*$                                      &$\frac{(z_*^2-4)}{2}$ &{\small $z_*^2-3 \gamma$} & $\frac{\left(z_*^2-6\right)}{2}$       & $-z_*^3 df_*$ \\ \hline
P$_{9}$       &$\sqrt{\frac{8}{3z_*^2}}$&$\frac{2}{\sqrt{3} z_*}$&$\frac{\sqrt{z_*^2-4}}{z_*}$ &$z_*$    &$b$ & \multicolumn{2}{c|}{$-\frac{1}{2}\pm\frac{\sqrt{64-15z_*^2}}{2 z_*}$} &$-4z_*df_*$ \\ \hline
P$_{10}$      &$\frac{\sqrt{6}\gamma}{2z_*}$&$\frac{\sqrt{6a\gamma}}{2z_*}$&0&$z_*$    &$-\frac{b}{2}$ &\multicolumn{2}{c|}{$-\frac{3a}{4}\pm\frac{3}{4}\frac{\sqrt{a(24\gamma^2-z_*^2(9\gamma -2))}}{z_*}$}  &$-3\gamma z_*df_*$
\\
\hline
\hline
\end{tabular}
\caption{Fixed points and the corresponding eigenvalues for the $\phi$-matter-radiation system without coupling, i.e., $Q=0$. Here $a = 2-\gamma$ and $b = 4 - 3\gamma$.}
\label{table1}
\end{table}

\paragraph{Properties of the fixed points}

The twelve fixed points are divided according to the classification of the potentials in \sect{classpot}. The points P$_{1-4}$ exist for all type of potentials while the points P$_{5-10}$ exist for only those potentials for which $f(z)$ vanishes for some finite $z_*$. The properties of these fixed points with their cosmological implications are as follows.

\paragraph{P$_1$} This point is the stable attractor solution when $z^2 f(z)|_{z=0}\geq0$ (stable spiral for $z^2 f(z)|_{z=0}>3/4$) and is also associated with the accelerated de Sitter phase with $w_{\mathrm{eff}} = -1$ and dominated by the scalar field with $\Omega_\phi = 1$. This can be an end-point or a future attractor (sink) and hence can give the observed late-time accelerated phase.

\paragraph{P$^{\pm}_{2,7}$} Of these, the points P$_2^\pm$ exist only for potentials for which $z^2 f(z)|_{z=0}=0$. All four points are dominated by the kinetic energy of the scalar field with $w_\phi = 1$ and $\Omega_\phi =1$. Note that for the points P$_2^\pm$, the eigenvalue $\lambda_3 = \kappa_1$ given by

\be
\kappa_1 =  \mp\sqrt{6}(2zf(z) + z^2\mathrm{d}f(z))|_{z=0}.
\ee

\noindent The nature of the eigenvalues for these points shows that these are the saddle points and hence can be associated with a transient phase.

\paragraph{P$_{3,5}$} These are matter (barotropic fluid) dominated points. This is reflected in the $w_{\mathrm{eff}}$ which is $\gamma -1$ and $\Omega_m =1$. These are also saddle points and hence again a transient phase where a phase trajectory can spend some time in the vicinity before moving towards a stable attractor.

\paragraph{P$_{4,6}$} These are the radiation dominated points with $w_{\mathrm{eff}} = 1/3$ and $\Omega_r =1$. These are also saddle points but given the initial conditions, these can be associated with the past attractors (or sources) since our universe was dominated by radiation in the early phase. The phase trajectories begin their evolution from this point.

\begin{table}[t!]
\centering
\begin{tabular}{|c|c|c|c|c|c|c|}
\hline\hline
Points                &Existence&Stability&$w_{\mathrm{eff}}$ &$w_\phi$ &$\Omega_\phi$&$\Omega_m$    \\
\hline\hline
P$_1$                &Always &$z^2 f(z)|_{z=0}>0$ &-1&-1&1&0                                                             \\ \hline
P$_2^\pm$        &''&Saddle&1&1&1&0                                                              \\ \hline
P$_3$                &''&''&$\gamma- 1$&Indet.&0&1                                                               \\ \hline
P$_4$                &''&''&$\frac{1}{3}$&"&0&0                                                                \\ \hline
P$_5$                &''&''&$\gamma -1$&"&0&1                                                                \\ \hline
P$_6$                &''&''&$\frac{1}{3}$&"&0&0                                                               \\ \hline
P$_7^\pm$        &''&''&1&1&1&0                                                                   \\ \hline
P$_8$                &$|z_*|\le\sqrt{6}~~$&See \eq{stability8zerobeta}&$\frac{1}{3} \left(z_*^2-3\right)$&$\frac{1}{3} \left(z_*^2-3\right)$&1&0          \\ \hline
P$_{9}$              &$|z_*|\geq 2~~$&See \eq{stability9zerobeta}&$\frac{1}{3}$&$\frac{1}{3}$&$\frac{4}{z_*^2}$&0                                                 \\ \hline
P$_{10}$            &$|z_*|\geq \sqrt{3\gamma }$& See \eq{stability10zerobeta} &$\gamma -1$&$\gamma -1$&$\frac{3 \gamma }{z_*^2}$&$1-\frac{3 \gamma }{z_*^2}$  \\ \hline
\hline
\end{tabular}
\caption{Existence and stability conditions for the fixed points for $Q=0$.}
\label{table2}
\end{table}

\paragraph{P$_{8-10}$} The properties of these points depend on the potentials since the value of $z_*$ is determined by the form of $f(z)$. These points can be stable under the following conditions:

\be
\label{stability8zerobeta}
\hspace{-3.45cm}\begin{array}{l}
\text{Stability}\\
\text{condition}\\
\text{for}~\mathrm{P}_{8}
\end{array}
\left\{
\begin{array}{ll}
1\leq\gamma \leq\frac{4}{3}   &\mathrm{and}\,\,-\sqrt{3\gamma}<z_*<\sqrt{3\gamma}\,\,{\mathrm{and}}\,\, z_*\mathrm{d}f_*>0 \\
\\
\frac{4}{3}\leq\gamma <2  &\mathrm{and}\,\,-2<z_*< 2 \,\,{\mathrm{and}}\,\, z_*\mathrm{d}f_*>0
\end{array}
\right.
\ee

\be
\label{stability9zerobeta}
\hspace{-0.265cm}\begin{array}{l}
\text{Stability}~
\text{condition}~
\text{for}~\mathrm{P}_{9}
\end{array}
\left\{
\begin{array}{ll}
\frac{4}{3}<\gamma <2  &\mathrm{and}\,\,(z_*< -2 \,\,\mathrm{or}\,\, z_*>2) \,\,\mathrm{and}\,\,z_*\mathrm{d}f_*>0
\end{array}
\right.
\ee

\be
\label{stability10zerobeta}
\begin{array}{l}
\text{Stability}~
\text{condition}\\
\text{for}~\mathrm{P}_{10}
\end{array}
\left\{
\begin{array}{ll}
1\leq\gamma <\frac{4}{3} &\left\{\begin{array}{ll} 3\gamma <z_*^2< \frac{24\gamma^2}{(9\gamma-2)}  \,\,\mathrm{and}\,\,z_*\mathrm{d}f_*>0  &\text{(Stable~node)}\\
z_*^2> \frac{24\gamma^2}{(9\gamma-2)} \,\,\mathrm{and}\,\,z_*\mathrm{d}f_*>0 &\text{(Stable~spiral)}.
\end{array}\right.
\end{array}
\right.
\ee

\noindent Point P$_8$ is dominated by the scalar field and exists for $z_*^2<6$. Taking $\gamma = 1$ for instance, that is, the barotropic fluid is a pressure-less dust, the point is stable for $z_*^2< 3$ and will give an accelerating phase for $z_*^2<2$ and $z_*\mathrm{d}f_*>0$. An interesting aspect of this point is that for $z_*^2 = 3$, the scalar field is able to mimic matter with $w_\phi = w_{\mathrm{eff}}= 0$. Point P$_9$ is usually \emph{absent} in the analysis done without having a separate radiation fluid component. This point offers radiation domination where the scalar field also behaves like radiation with $w_\phi  = 1/3$. Point P$_{10}$ is the scaling solution (with $\rho_\phi/\rho_m = \text{const}$) where neither of the components completely dominates the evolutionary dynamics. However, at this point, we have $w_{\mathrm{eff}} = w_\phi = \gamma - 1$. So, an accelerating phase is not possible with $\gamma\geq 1$. Also, it is to be noted that the stability conditions for these points are mutually exclusive so that these cannot be stable together unless in a very specific case.

The above analysis of the critical points shows that we can obtain an accelerated expansion provided that the solutions approach the fixed point P$_1$ or P$_8$ in which case the final state of the universe is the scalar-field dominated one $(\Omega_\phi = 1)$. The scaling solutions P$_9$ and  P$_{10}$ are not viable to explain the late-time acceleration. However, they can be used to provide the cosmological evolution in which the scalar field scales proportionally to radiation and matter.

\subsection{Conformal coupling between the scalar field and matter fluid}

\begin{table}[t!]
\centering
\begin{tabular}{|c|c|c|c|c|c|c|c|c|}
\hline\hline
Points   &$x$ &$y$ &$u$ &$z$         &$\lambda_1$ &$\lambda_2$&$\lambda_3$&$\lambda_4$ \\
\hline\hline
P$_1$         & 0  &1  &0  &0        &-2 &$-3\gamma$ &\multicolumn{2}{c|}{{\small $-\frac{3}{2}\pm\frac{3}{2}\sqrt{1-\frac{4}{3}z^2f(z)}|_{0}$}}\\ \hline
P$_2^\pm$   & $\pm1$ &0 &0 &0        &3 &1&{\small $3a\mp2\beta$}&$\kappa_1$\\ \hline
P$_3$         &$\frac{b}{2\beta}$ &0 &$\sqrt{1-\frac{3ab}{4\beta^2}}$ &0      &2 &$\kappa_2$ &\multicolumn{2}{|c|}{{\small $-\frac{1}{2} \pm \frac{\sqrt{3}}{2\beta}\sqrt{c\beta^2 + a b^2 }$}} \\ \hline
P$_4$        &$\frac{2\beta}{3a}$ &0 &0 &0    &$\frac{4\beta ^2-9a^2}{6a}$ &$\frac{4\beta ^2 - 3 ab}{6a}$ &$\frac{4\beta ^2+9\gamma a}{6a}$ &$\kappa_3$\\ \hline
P$_5$         & 0 &0 &1 &$z_a$     &0 &2 &-1 &$b$\\ \hline
P$_6^\pm$  &$\pm1$ &0 &0 &$z_*$        &1 &{\small $3a\mp2\beta$}&$\frac{6\mp\sqrt{6}z_*}{2}$ &{\small $\mp\sqrt{6}z_*^2\mathrm{d}f_*$}\\ \hline
P$_7$      &$\sqrt{\frac{8}{3z_*^2}}$ &$\frac{2}{z_*\sqrt{3}}$ &$\sqrt{1-\frac{4}{z_*^2}}$ &$z_*$  &$\kappa_4$ &{\small $-4z_*\mathrm{d}f_*$} &\multicolumn{2}{c|}{{\small $ -\frac{1}{2}\pm\sqrt{\frac{16}{z_*^2}-\frac{15}{4}}$}} \\ \hline
P$_8$          &$\frac{z_*}{\sqrt{6}}$ &$\sqrt{1- \frac{z_*^2}{6}}$ &0 &$z_*$     &$\frac{z_*^2 -6}{2}$ &$\frac{z_*^2 -4}{2}$ &$\kappa_5$ &{\small $-z_*^3\mathrm{d}f_*$}\\  \hline
P$_{9}$      &$\frac{b}{2\beta}$ &0 &$\sqrt{1-\frac{3ab}{4\beta^2}}$ &$z_*$         &$\frac{8\beta-b\sqrt{6}z_*}{4\beta}$ &$\kappa_6$ &\multicolumn{2}{c|}{{\small $-\frac{1}{2} \pm \frac{\sqrt{3}}{2\beta}\sqrt{c\beta^2 + a b^2 }$}} \\ \hline
P$_{10}$        &$\frac{2\beta}{3a}$ &0 &0 &$z_*$      &$\frac{4\beta ^2-9a^2}{6a}$&$\frac{4\beta ^2 - 3 ab}{6a}$ &$\kappa_7$ &$\kappa_8$\\ \hline
P$_{11}$      &$x_{11}$ &$y_{11}$  &0 &$z_*$   &$\kappa_9$ &$\kappa_{10}$ &\multicolumn{2}{c|}{$\frac{-s\pm\sqrt{s^2-4t}}{4(\sqrt{6} z_*-2 \beta)^2}$}\\
\hline\hline
\end{tabular}
\caption{Fixed points and the corresponding eigenvalues for $Q=\sqrt{2/3}\kappa\beta\rho_m\dot{\phi}$ where the coupling parameter $\beta$ is assumed to be constant. Here $a= 2-\gamma$, $b = 4-3\gamma$ and $c=4\gamma-5$. The point P$_{11}$ has $x_{11} = 3\gamma/(\sqrt{6}z_*-2\beta)$ and $y_{11} = \sqrt{4\beta^2 +9a \gamma - 2\sqrt{6}\beta z_*}/(\sqrt{6}z_*-2\beta)$.}
\label{table3}
\end{table}
Having studied the dynamics of the scalar field without any interaction with the matter fluid, we introduce a coupling between the two of the form, $Q =\sqrt{2/3}\kappa\beta\rho_m\dot{\phi}$ in this section. This form of the coupling can be motivated \cite{cup1,cup2} by considering a coupled gravity and scalar field as a scalar-tensor theory in the Jordan frame and then (conformally) transforming to Einstein frame giving a coupled quintessence in normal Einstein gravity. In this way the scalar field gets coupled to all matter except radiation for which the stress-energy tensor is traceless. The coupling parameter $\beta$ can be a constant or time-dependent through an explicit $\phi$ dependence. We shall consider a constant $\beta$ for our analysis and comment about the variable coupling later. The presence of coupling adds another degree of freedom in the parameter space of the quintessence models. Thus, we see the existence of new fixed points over and above those present in the analysis without any interaction and also modifications to some of the previous points.

\paragraph{Properties of the fixed points} We have points P$_3$, P$_4$, P$_9$ and P$_{10}$ as the new additions and modification of the scaling point P$_{10}$ (in Table~\ref{table1}) to Point P$_{11}$ in the new list (Tables~\ref{table3} and \ref{table4}) with the inclusion of coupling. For some of the previous points that still exist with coupling, the eigenvalues get modified with the interaction coming in affecting the stability conditions with another parameter.

\paragraph{P$_{3,9}$} These are the new points that occur in the presence of coupling between the scalar field and matter fluid. The condition for their existence are
\be
\label{exist3}
1\leq \gamma \leq4/3\quad\text{and}\quad |\beta| \geq \sqrt{3}\sqrt{3 \gamma ^2-10 \gamma +8}/2.
\ee
While P$_3$ is always a saddle point, P$_9$ can be stable under following conditions:

\begin{table}[t!]
\centering
\begin{tabular}{|c|c|c|c|c|c|c|}
\hline\hline
Points   &Existence  &Stability  &$w_{\mathrm{eff}}$  &$w_\phi$ &$\Omega_\phi$  &$\Omega_m$\\
\hline\hline
P$_1$    & $\forall \beta$ &$z^2 f(z)|_{z=0}>0$ &-1 &-1&1&0     \\ \hline
P$_2^\pm$   & $\forall \beta$ &Saddle$/$Unstable &1 &1&1&0\\ \hline
P$_3$         &\eq{exist3}  &Always Saddle &$\frac{1}{3}$ &1  &$\frac{b^2}{4\beta^2}$& $\frac{b}{2\beta^2}$ \\ \hline
P$_4$      &$|\beta| \leq 3(2-\gamma)/2$   &Saddle$/$Unstable & {\small $\gamma-1 +\frac{4\beta^2}{9a}$} &1 &$\frac{4\beta^2}{9a^2}$ &$1-\frac{4\beta^2}{9a^2}$ \\ \hline
P$_5$        &$\forall \beta$ &Always Saddle &$\frac{1}{3}$ &Indet. &0 &0  \\ \hline
P$_6^\pm$  &$\forall \beta$ &Saddle$/$Unstable & 1&1 &1&0 \\ \hline
P$_7$          &$|z_*|\geq2$  & \eq{stability7}  &$\frac{1}{3}$ &$\frac{1}{3}$   &$\frac{4}{z_*^2}$&0\\ \hline
P$_8$          & $|z_*|\leq\sqrt{6}$   & \eq{stability8}   & $\frac{z_*^2}{3}-1$ &$\frac{z_*^2}{3}-1$ &1&0  \\  \hline
P$_{9}$      &\eq{exist3}&\eq{stability9}&$\frac{1}{3}$& 1     &$\frac{b^2}{4\beta^2}$&  $\frac{b}{2\beta^2}$ \\ \hline
P$_{10}$      &$|\beta| \leq 3(2-\gamma)/2$  &\eq{stability10} &  {\small $\gamma -1 +\frac{4\beta^2}{9a}$} &1  &$\frac{4\beta^2}{9a^2}$ & $1-\frac{4\beta^2}{9a^2}$  \\ \hline
P$_{11}$     &\eq{exist11}   &\fig{stability11} &\eq{weff11} &\multicolumn{2}{c|}{\eq{param11}}   &{\small $1- \Omega_\phi$} \\
\hline\hline
\end{tabular}
\label{interaction1}
\caption{Existence and stability conditions for the fixed points with coupling given by $Q=\sqrt{2/3}\kappa\beta\rho_m\dot{\phi}$ where the coupling parameter $\beta$ is assumed to be constant.}
\label{table4}
\end{table}

\be
\label{stability9}
\begin{array}{l}
\text{Stability}\\
\text{condition}\\
\text{for}~\mathrm{P}_{9}
\end{array}
\left\{
\begin{array}{ll}
1\leq\gamma <\frac{5}{4} &\left\{\begin{array}{l}-\sqrt{\frac{9 \gamma ^3-42 \gamma ^2+64 \gamma -32}{4 \gamma -5}}<\beta <-\frac{\sqrt{3}}{2}  \sqrt{3 \gamma ^2-10 \gamma +8}\\
\quad\text{and}\quad z_*<-\frac{8 \beta }{3 \sqrt{6} \gamma -4 \sqrt{6}}\,\,\text{and}\,\,\text{d}f_* < 0\\
\\
\frac{\sqrt{3}}{2} \sqrt{3 \gamma ^2-10 \gamma +8}<\beta <\sqrt{\frac{9 \gamma ^3-42 \gamma ^2+64 \gamma -32}{4 \gamma -5}}\\
\quad\text{and} \quad z_*>-\frac{8 \beta }{3 \sqrt{6} \gamma -4 \sqrt{6}} \,\,\text{and}\,\,\text{d}f_* > 0 \end{array}\right.\\
\\
\frac{5}{4}\leq\gamma <\frac{4}{3} &\,\,\left\{\begin{array}{l}\beta <-\frac{\sqrt{3}}{2}\sqrt{3 \gamma ^2-10 \gamma +8}\\
\text{and}\,\,z_*<-\frac{8 \beta }{3 \sqrt{6} \gamma -4 \sqrt{6}}\,\,\text{and}\,\,\text{d}f_* < 0\\
\\
 \beta >\frac{\sqrt{3}}{2} \sqrt{3 \gamma ^2-10 \gamma +8}\\
 \text{and}\,\,z_*>-\frac{8 \beta }{3 \sqrt{6} \gamma -4 \sqrt{6}}\,\,\text{and}\,\,\text{d}f_* > 0
\end{array}
\right.
\end{array}
\right.
\ee
Note the eigenvalues referred to in the Table~\ref{table3} for these points have
\be
\kappa_2= -\sqrt{\frac{3}{2}}\,\frac{(4-3\gamma)}{\beta}(2zf(z) + z^2\mathrm{d}f(z))\Big|_{z=0}; \hspace{10pt} \kappa_6 = -\sqrt{\frac{3}{2}}\frac{(4-3\gamma)z_*^2\mathrm{d}f_*}{\beta}.
\ee
Both the points have an effective radiation dominated behaviour with $w_{\mathrm{eff}} = 1/3$ but have $w_\phi = 1$. These points are associated with scaling behaviour which can be stable in the case of P$_9$ as shown above.

\paragraph{P$_{4,10}$} These are also scaling points with $\Omega_r =0$ and $\Omega_\phi/\Omega_m \sim \mathrm{constant}$ for constant $\beta$. The point P$_4$ is saddle or unstable while P$_{10}$ can be stable under the following conditions:
\be
\label{stability10}
\begin{array}{l}
\text{Stability}\\
\text{condition}\\
\text{for}~\mathrm{P}_{10}
\end{array}
\begin{array}{ll}
\quad1\leq\gamma <\frac{4}{3} &\left\{\begin{array}{l}-\frac{\sqrt{3}}{2}\sqrt{3 \gamma ^2-10 \gamma +8}<\beta <0\\
\quad\text{and}\quad z_*<\frac{4 \beta ^2-9 \gamma ^2+18 \gamma }{2 \sqrt{6} \beta}\,\,\text{and}\,\,\text{d}f_* < 0\\
\\
0<\beta <\frac{\sqrt{3}}{2}\sqrt{3\gamma ^2-10 \gamma +8}\\
\quad\text{and}\quad z_*>\frac{4 \beta ^2-9 \gamma ^2+18 \gamma }{2 \sqrt{6} \beta}\,\,\text{and}\,\,\text{d}f_* > 0\
\end{array}
\right.
\end{array}
\ee

The eigenvalues of these points as given in Table~\ref{table3} are given by
\begin{align}
&\kappa_3 =  -\sqrt{\frac{2}{3}}\frac{2\beta}{(2-\gamma)}(2zf(z) + z^2\mathrm{d}f(z))\Big|_{z=0}\nonumber\\
&\kappa_7 = \frac{9\gamma(2-\gamma) +4 \beta ^2-2 \sqrt{6} \beta  z_*}{6 (2-\gamma)};\hspace{10pt} \kappa_8 = -\sqrt{\frac{2}{3}}\frac{2\beta z_*^2\mathrm{d}f_*}{2-\gamma}.
\end{align}
For certain values of $\beta$, these points can give accelerated expansion with effective equation of state parameter given by $w_{\rm{eff}} = \gamma -1 + 4\beta^2/9(2-\gamma)$.

\paragraph{P$_{7,8}$} These points are same as points P$_{8,9}$ in the analysis without coupling. With the inclusion of $\beta$, the stability conditions for these points get modified to

\be
\label{stability7}
\begin{array}{l}
\text{Stability}\\
\text{condition}\\
\text{for}~\mathrm{P}_{7}
\end{array}
\left\{
\begin{array}{ll}
1\leq \gamma <\frac{4}{3} & \left\{\begin{array}{lll}\beta <-\frac{12-9 \gamma}{2 \sqrt{6}} &\quad\text{and}\quad \frac{4 \sqrt{6} \beta}{12-9\gamma}<z_*<-2 &\,\,\text{and}\,\,\text{d}f_* < 0\\ \beta >\frac{12-9 \gamma }{2 \sqrt{6}}&\quad\text{and} \quad 2<z_*<\frac{4 \sqrt{6} \beta}{12-9 \gamma} &\,\,\text{and}\,\,\text{d}f_* > 0 \end{array}\right.  \\
\\
\gamma = \frac{4}{3}& \left\{\begin{array}{lll}\beta < 0 &\quad\text{and}\quad z_*<-2&\,\,\text{and}\,\,\text{d}f_* < 0\\ \beta >0 &\quad\text{and}\quad z_*>2&\,\,\text{and}\,\,\text{d}f_* > 0\end{array}\right. \\
\\
\frac{4}{3}<\gamma <2 & \left\{\begin{array}{ll}
\beta <-\frac{9 \gamma-12}{2 \sqrt{6}}&\text{and}\quad z_*<-2 \,\,\text{or}\,\, z_*>-\frac{4 \sqrt{6} \beta }{9 \gamma-12}\\
-\frac{9 \gamma-12 }{2 \sqrt{6}}\leq \beta \leq \frac{9 \gamma -12}{2 \sqrt{6}} &\text{and}\quad z_*<-2\,\,\text{or}\,\,z_*>2\\
\beta >\frac{9 \gamma-12}{2 \sqrt{6}}&\text{and}\quad z_*<-\frac{4 \sqrt{6} \beta }{9 \gamma -12}\,\,\text{or}\,\, z_*>2\end{array}\right\}z_*\text{d}f_*>0
\end{array}
\right.
\ee

\be
\label{stability8}
\begin{array}{l}
\text{Stability}\\
\text{condition}\\
\text{for}~\mathrm{P}_{8}
\end{array}
\left\{
\begin{array}{ll}
1\leq\gamma <2 \,\,\text{and}\,\, \text{d}f_*<0  &\,\,\left\{\begin{array}{ll}\beta\leq\frac{9 \gamma-12}{2 \sqrt{6}} &\quad\text{and}\quad -2<z_*<0\\ \beta >\frac{9 \gamma-12}{2 \sqrt{6}}&\quad\text{and} \quad \frac{\beta-\sqrt{\beta ^2+18 \gamma}}{\sqrt{6}} <z_*<0 \end{array}\right.  \\
\\
1\leq\gamma <2 \,\,\text{and}\,\, \text{d}f_*>0  &\,\,\left\{\begin{array}{ll}\beta<\frac{12-9 \gamma}{2 \sqrt{6}} &\quad\text{and}\quad 0<z_*<\frac{\beta+\sqrt{\beta ^2+18 \gamma}}{\sqrt{6}}  \\ \beta \geq\frac{12-9 \gamma}{2 \sqrt{6}}&\quad\text{and} \quad 0 <z_*<2 \end{array}\right.
\end{array}
\right.
\ee
The eigenvalues $\kappa_4$ and $\kappa_5$ for these points are
\be
\kappa_4 = \frac{-4 \sqrt{6} \beta  z_*-9 \gamma  z_*^2+12 z_*^2}{3 z_*^2} ;\hspace{10pt}
\kappa_5 = -3 \gamma -\sqrt{\frac{2}{3}} \beta  z_*+z_*^2;
\ee
All the properties for these points are same as that of points P$_{8,9}$ without coupling.

\paragraph{P$_{11}$} The scaling point P$_{10}$ gets modified in the presence of coupling and we designate it as P$_{11}$ in the new table. With non-zero $\beta$, the conditions for existence of this point are

\begin{align}
\label{exist11}
&\begin{array}{l}
\text{Existence}\\
\text{condition}\\
\text{for}~\mathrm{P}_{11}
\end{array}
\left\{
\begin{array}{ll}
\beta = 0 & \quad |z_*|\geq \sqrt{3\gamma}\\
0<\beta<\frac{3}{2}(2-\gamma)& \quad z_*\leq \frac{\beta-\sqrt{\beta ^2+18 \gamma }}{\sqrt{6}}\quad\mathrm{or}\quad \frac{\beta+\sqrt{\beta ^2+18 \gamma }}{\sqrt{6}}\leq z_*\leq \frac{4 \beta ^2-9 \gamma ^2+18 \gamma }{2 \sqrt{6} \beta }\\
\beta > \frac{3}{2}(2-\gamma)& \quad z_*\leq \frac{\beta-\sqrt{\beta ^2+18 \gamma }}{\sqrt{6}}
\end{array}
\right.
\end{align}
for $\beta>0$. The signs in the above expressions change accordingly for $\beta<0$. The parameters $w_{\rm eff}$, $w_\phi$, $\Omega_\phi$ and $\Omega_m$ for this point are
\be
\label{weff11}
w_{\rm eff} = (\gamma -1)+\frac{2 \beta  \gamma }{\sqrt{6} z_*-2 \beta };
\ee
\be
\label{param11}
w_\phi = -\frac{9 \gamma ^2}{-2 \beta ^2-9 \gamma +\sqrt{6} \beta  z_*}-1;\hspace{5pt}\Omega_\phi = \frac{4 \beta ^2+18 \gamma -2 \sqrt{6} \beta  z_*}{\left(\sqrt{6} z_*-2 \beta \right)^2};
\hspace{5pt}\Omega_m = 1 - \Omega_\phi
\ee
\begin{figure}[t!]
\centering
\includegraphics[width=0.45\textwidth,scale=0.2]{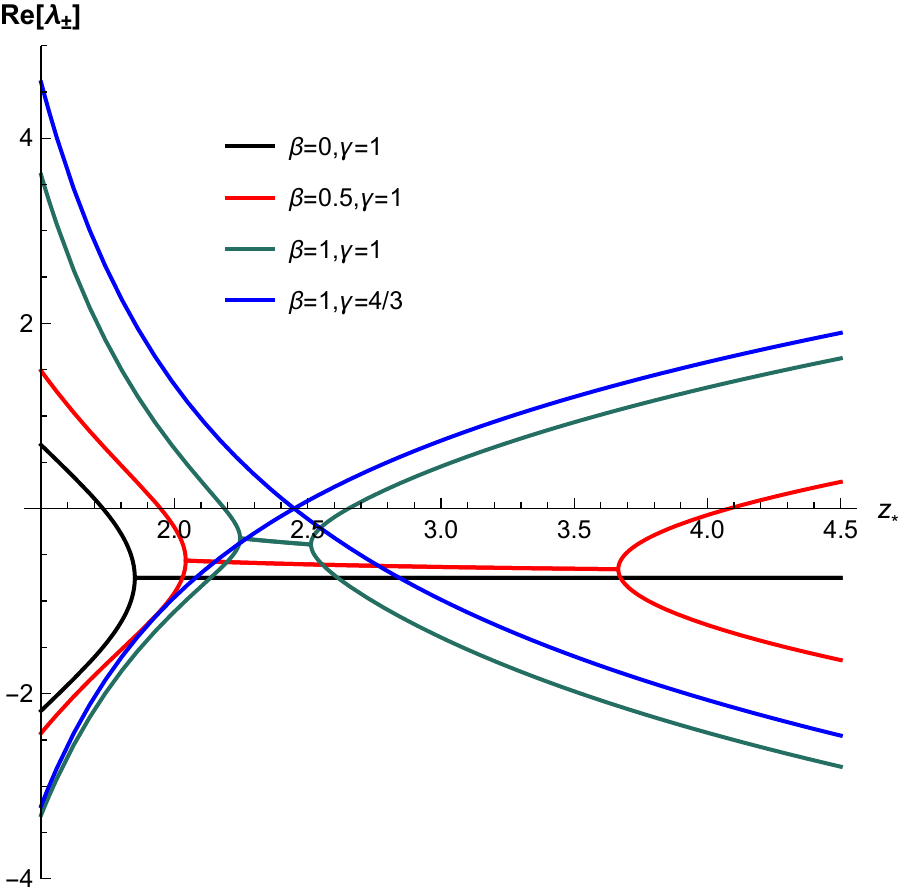}\hfill
\includegraphics[width=0.45\textwidth,scale=0.2]{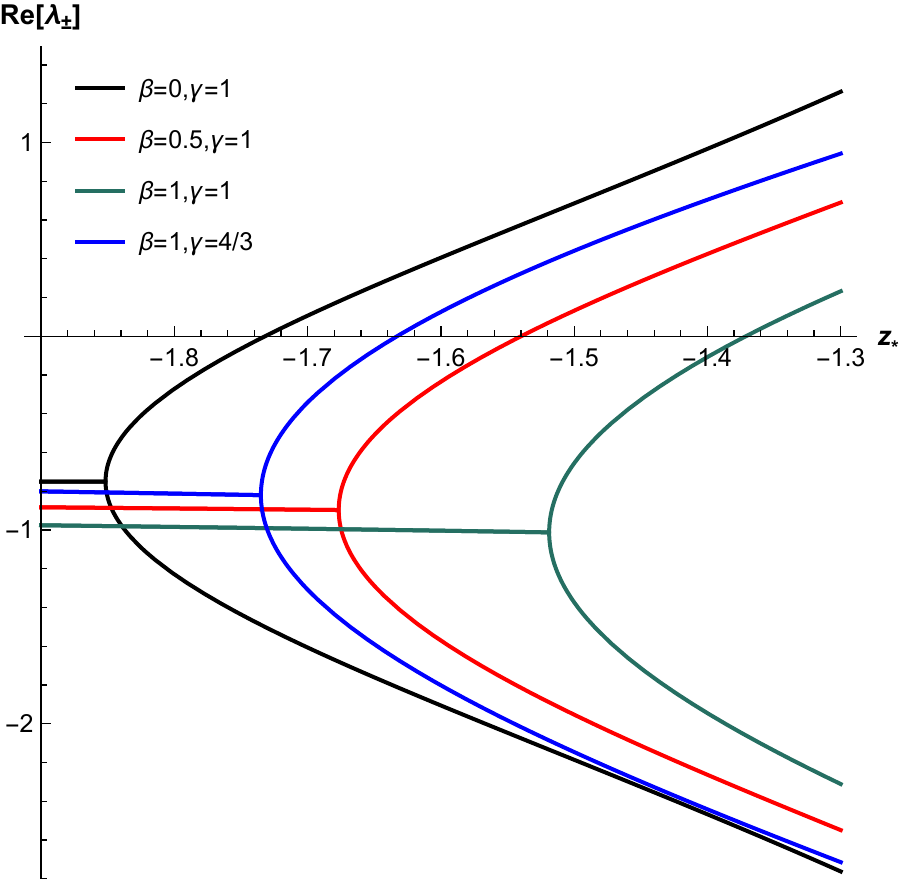}
\caption{Stability for the point P$_{11}$ can be assessed graphically from the real part of the eigenvalues $\lambda_\pm$ if the other two eigenvalues are negative for the same set of parameter values. The plots here show the forking diagrams of the eigenvalues for ranges in $z_*<0$ (left) and $z_*>0$ (right) on respectively. The point will be stable for given $\beta$ and $\gamma$ for the range in $z_*$ for which the real parts of $\lambda_\pm$ (the forks here) lie below zero. When the real part is quasi-constant and un-forked, the eigenvalues are complex. In this case, if the real part is negative, the point will be a stable spiral.}
\label{stability11}
\end{figure}

\noindent The complicated nature of the eigenvalues:
\be
\kappa_9 = \gamma  \left(\frac{3 \beta }{\sqrt{6} z_*-2 \beta }+\frac{3}{2}\right)-2; \hspace{10pt}\kappa_{10} = -\frac{3 \sqrt{6} \gamma z_*^2\mathrm{d}f_*}{\sqrt{6} z_*-2 \beta};\hspace{10pt} \lambda_\pm = \frac{-s\pm\sqrt{s^2-4t}} {4(\sqrt{6} z_*-2 \beta)^2}
\ee
where $s$ and $t$ are given by
\be
s = 18 a z_*^2+24 \beta ^2+6 \sqrt{6} \beta  \gamma z_*-24 \sqrt{6} \beta z_*
\ee
and
\begin{align}
t =& -1296 a^2 \beta ^2 \gamma ^2-972 a^2 \gamma ^2 z_*^2+1296 \sqrt{6} a^2 \beta  \gamma ^2 z_*-576 a \beta ^4 \gamma -1296 a \beta ^2 \gamma ^3+1296 a \beta ^2 \gamma ^2\nonumber\\
&+648 a \gamma  z_*^4-432 \sqrt{6} a \beta  \gamma  z_*^3-864 a \beta ^2 \gamma  z_*^2-972 a \gamma ^3 z_*^2+720 \sqrt{6} a \beta ^3 \gamma  z_*+1296 \sqrt{6} a \beta  \gamma ^3 z_*\nonumber\\
&-1296 \sqrt{6} a \beta  \gamma ^2 z_* -576 \beta ^4 \gamma ^2+576 \beta ^4 \gamma -144 \sqrt{6} \beta  z_*^5+1152 \beta ^2 z_*^4-576 \sqrt{6} \beta ^3 z_*^3\nonumber\\
&+216 \sqrt{6} \beta  \gamma ^2 z_*^3+768 \beta ^4 z_*^2-2160 \beta ^2 \gamma ^2 z_*^2+1728 \beta ^2 \gamma  z_*^2-64 \sqrt{6} \beta ^5 z_*+864 \sqrt{6} \beta ^3 \gamma ^2 z_*\nonumber\\
&-864 \sqrt{6} \beta ^3 \gamma  z_*
\end{align}
makes it difficult to assess the conditions for stability for this point analytically. The point will be stable node if all the eigenvalues are negative and a stable spiral if $\lambda_\pm$ are complex with the real part being negative. We depict this for some of the cases in the Fig.~\ref{stability11}. The plots show the forking diagrams of the eigenvalues for ranges in $z_*<0$ (left) and $z_*>0$ (right) on respectively. The point will be stable for given $\beta$ and $\gamma$ for the range in $z_*$ for which the real parts of $\lambda_\pm$ (the forks here) lie below zero. When the real part is quasi-constant and un-forked, the eigenvalues are complex. In this case, if the real part is negative, the point will be a stable spiral.

We assumed the coupling parameter $\beta$ to be constant but generally it is $\phi$ dependent or time-dependent. In this case, we need another equation to close the autonomous system. Defining another variable $s = \kappa \phi$, this gives $ds/dN = \sqrt{6} x$. For the fixed points of the new system, we then need the right hand side of this equation to vanish as well. This makes $x=0$ necessary for all the fixed points. Then, the subset of fixed points listed in Table~\ref{table3} for which $x=0$ form the instantaneous fixed points for the system with changing $\beta$. The exact analysis would then require the functional form of the variable coupling to assess the exact dynamics.

\paragraph{Interpretation of the fixed-point analysis.} Before we proceed further, we shall comment briefly about the methodology used and the results obtained in this section. As noted earlier in the previous sections, the cosmological system presents a complicated non-linear problem in which the dynamics of the scale factor depends on the constituent fluids in a non-trivial manner. The inclusion of a scalar field to provide a dynamical dark energy constituent further adds on to the complexity of the problem. Then, to ascertain the dynamics of the system, we need to solve simultaneously, the Friedmann equation and the Klein-Gordon equation of the scalar field which gives the evolution of the scale factor and the scalar field given the initial conditions and the potential of the scalar field. The system is thus far from being solvable analytically, in general, and we need to resort to numerical techniques.

However, the otherwise analytically opaque dynamics can still be understood, albeit, in the manner of computing the fixed points of the non-linear equations. This is a fairly standard technique in non-linear dynamics which has been employed in dark energy models as well (see, for example, ref.~\cite{ed}). The fixed points are to be taken as the stages a non-linear system may go through during its evolution given the initial conditions. We adopted a similar procedure in this section with some differences and additions, viz., inclusion of radiation fluid, keeping the matter fluid as a barotropic one (with non-zero pressure), form of the potential for a scalar field to be generic and, including a coupling between the scalar field and matter sector. This, in a way, stretches the scheme of things to be as generic and tractable as possible. Thus, the fixed points obtained in this section form a super-set of all possible stages that a $\phi$CDM model (with and without coupling) can go through during its evolution. Then, for a specific potential of the scalar field, the allowed set of fixed points for that potential is a subset of this super-set and these fixed points give an instantaneous behaviour of the system in the phase space. Another aspect to be understood is that, although given a potential we can have a set of fixed points for the same, the points a system will traverse on its trajectory in the phase space depends on the choice of the initial conditions. Thus, the set of fixed points and their nature gives us a possible set of trajectories in the phase space which in itself is worth knowing, given that we can hardly infer anything analytically otherwise. The estimation of parameters and testing the concordance of models with the observations, in the end, is a numerical exercise.      

\begin{figure*}[ht!]
\centering
\includegraphics[width=0.45\textwidth,scale=0.2]{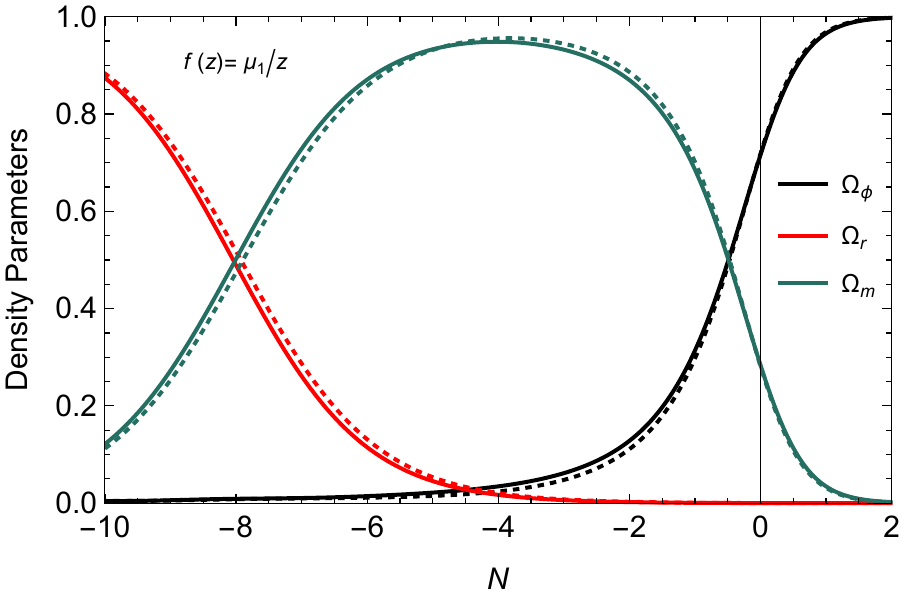}
\includegraphics[width=0.45\textwidth,scale=0.2]{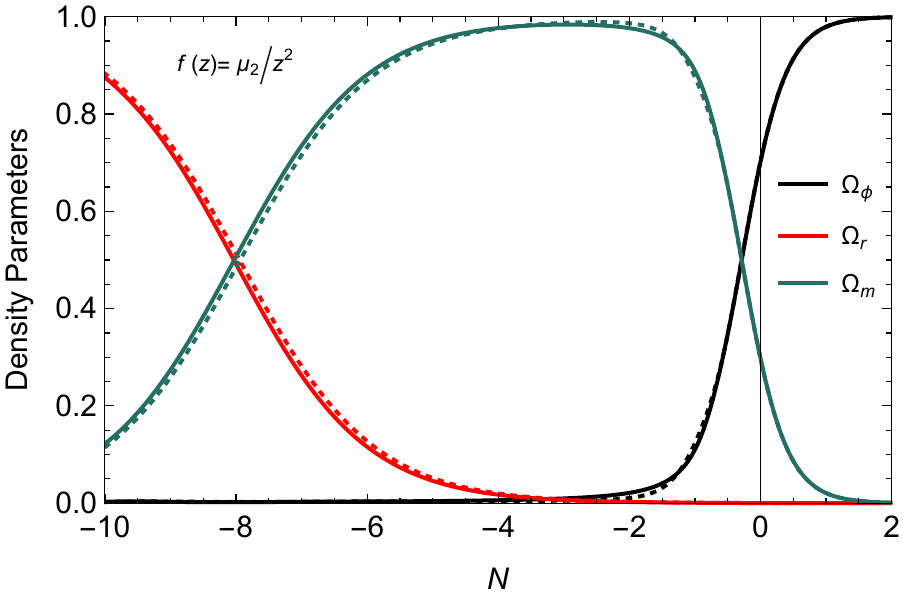}
\includegraphics[width=0.45\textwidth,scale=0.2]{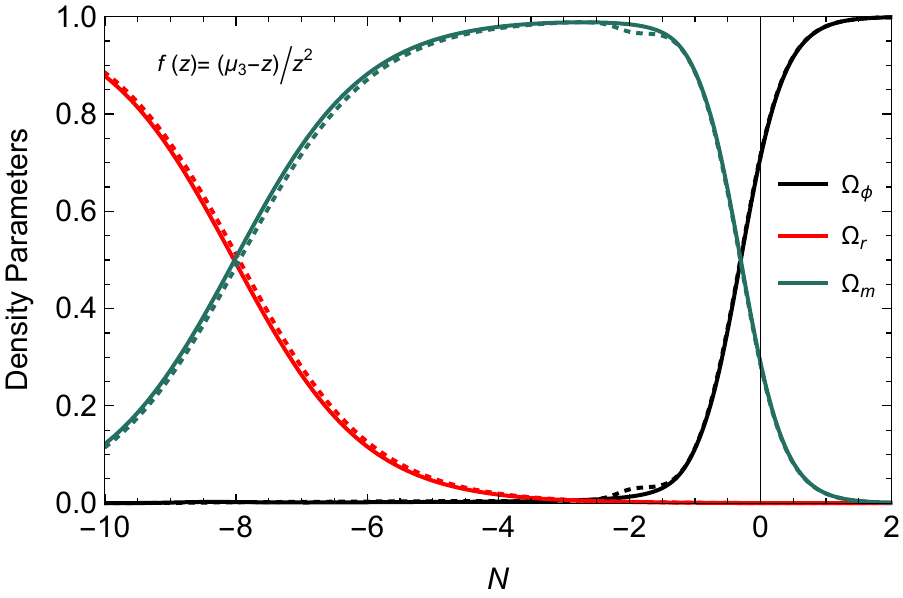}
\includegraphics[width=0.45\textwidth,scale=0.2]{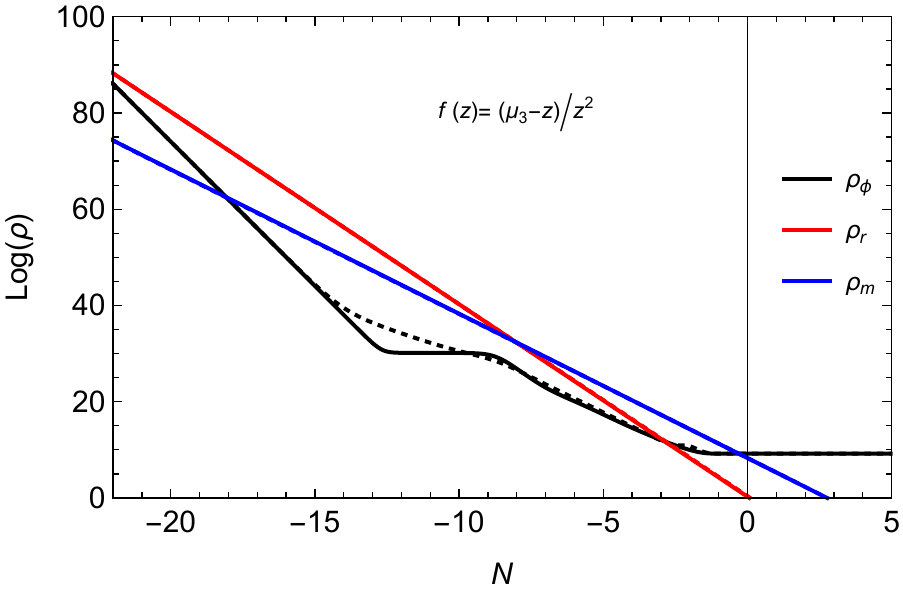}
\includegraphics[width=0.45\textwidth,scale=0.2]{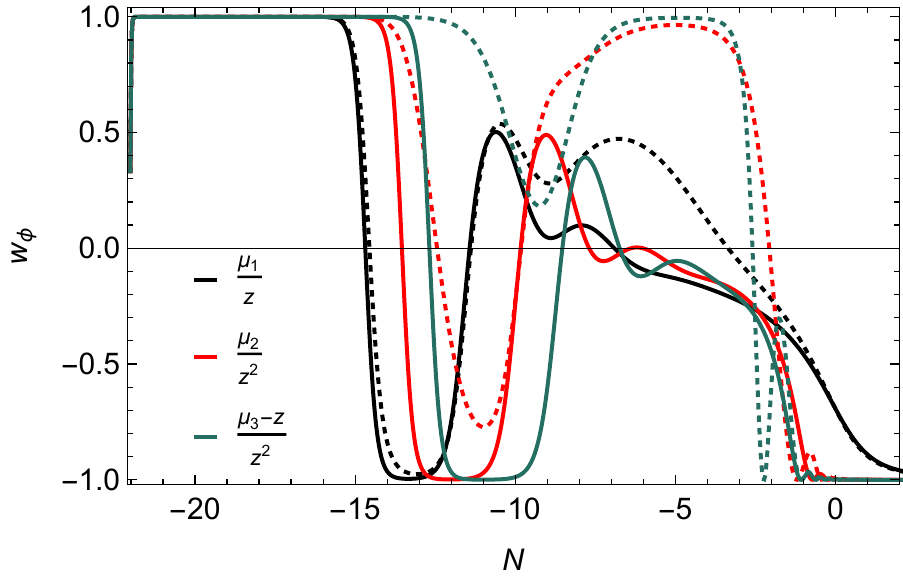}
\includegraphics[width=0.45\textwidth,scale=0.2]{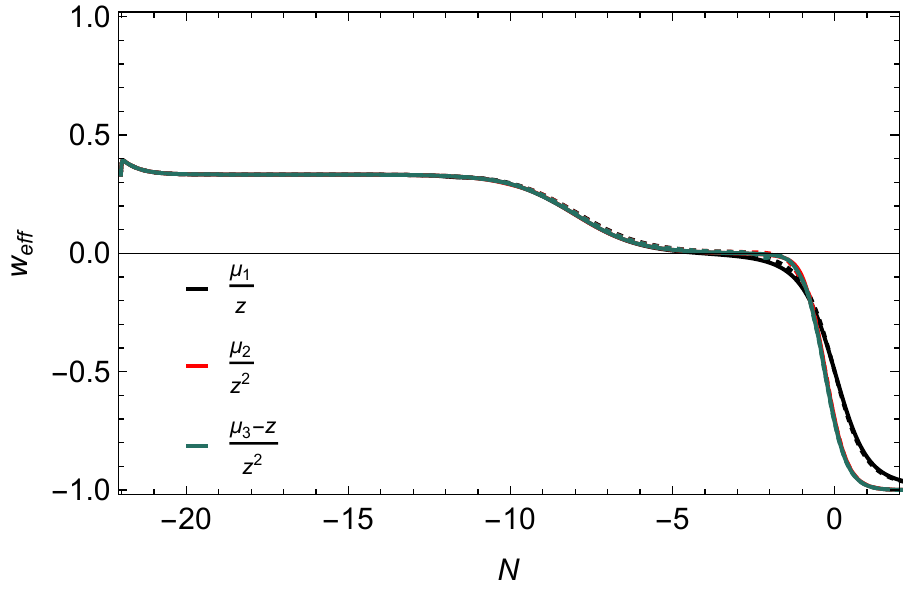}
\caption{{\bf Numerical peek at the cosmological dynamics I.} The plots show the evolution of the density parameters (top row and middle-left) for $f(z)$ given by $\mu_1/z$, $\mu_2/z^2$ and $(\mu_3-z)/z^2$ with (dashed) and without (regular) coupling. The parameters are $\mu_1 = 1$, $\mu_2=39$ and $\mu_3 = 100$ and the coupling constant $\beta = 0.1$. All three cases show tracker behaviour for the scalar field. This is particularly evident in the plot showing the evolution of density (middle-right) for $f(z) = (\mu_3-z)/z^2$. Finally, we show the evolution of $w_\phi$ and $w_{\rm eff}$ (bottom row) for the three cases.}
\label{plots1}
\end{figure*}

\section{A numerical peek at the cosmological dynamics}
\label{numerics}

Having analysed the radiation-matter-quintessence dynamical system using the linear stability analysis, we now look at the exact (using numerics) dynamics for a few examples. We take the function $f(z)$ to be of the form: $\mu/z^2$, $\mu/z$ and $(\mu-z)/z^2$ with and without coupling and solve the dynamical equations for the observed cosmological parameters. These pertain to the potentials $V(\phi)$ of the form: $V_0 \exp(-\mu\phi^2/2M_p^2)$, $V_0 \exp(\exp(-\mu \phi/M_p))$ and $V_0 \exp(-\exp(\phi/M_P)-\mu\phi/M_P)$ respectively. The first two cases do not have a finite $z_*$ and the third case falls under the category $z_*=\mu$ with $z^2f(z)|_{z=0} \neq 0$. Finally, we also consider a case $f(z)=(\mu-z)$ which has $z_*=\mu$ and $z^2f(z)|_{z=0} = 0$. This suffices to illustrate and explain the behaviour of models in accordance with the fixed point analysis of the previous section. We also assume that the barotropic fluid is perfect matter, that is, we take $\gamma =1$ for all the analysis henceforth.

\begin{figure*}[ht!]
\centering
\includegraphics[width=0.45\textwidth,scale=0.2]{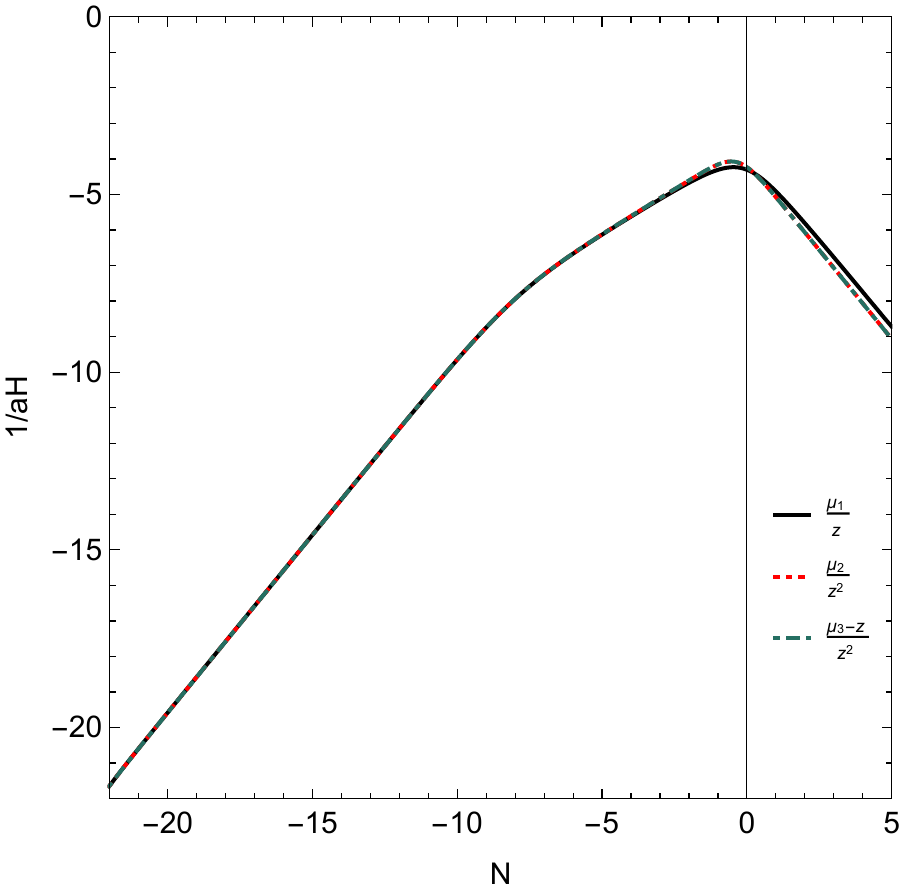}
\includegraphics[width=0.45\textwidth,scale=0.2]{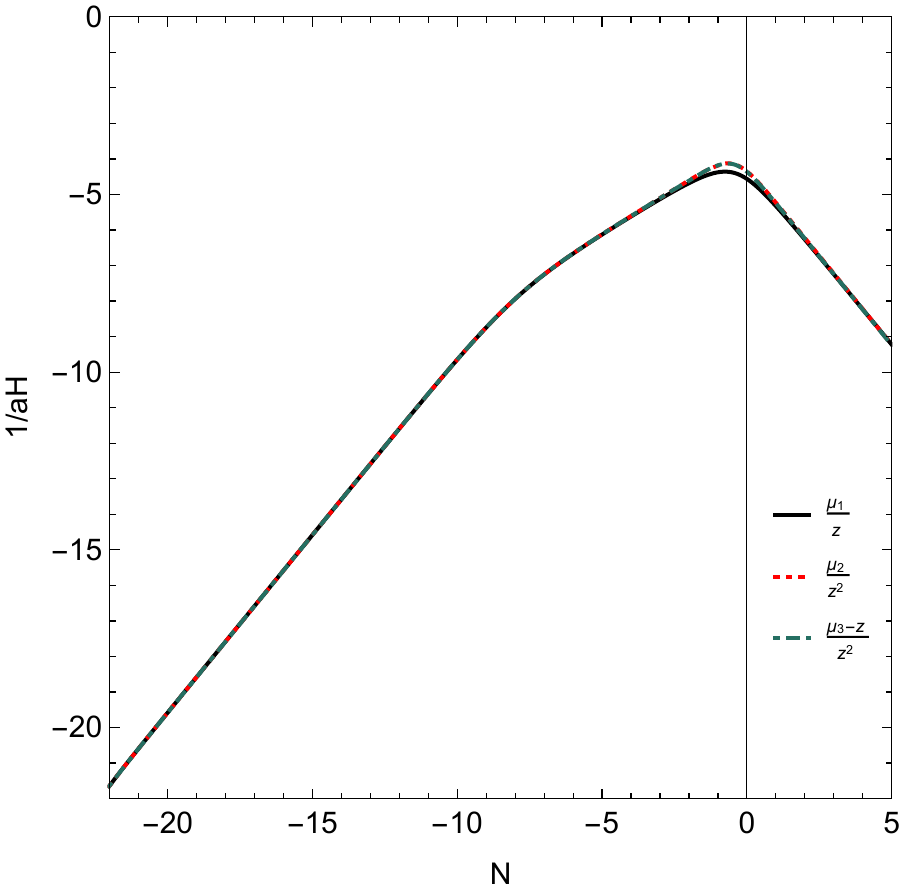}
\includegraphics[width=0.45\textwidth,scale=0.2]{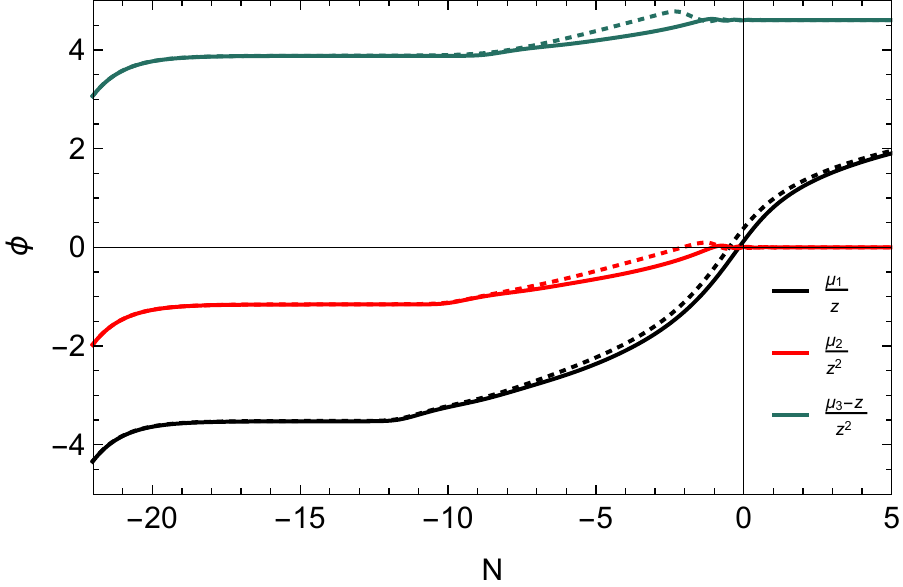}
\includegraphics[width=0.45\textwidth,scale=0.2]{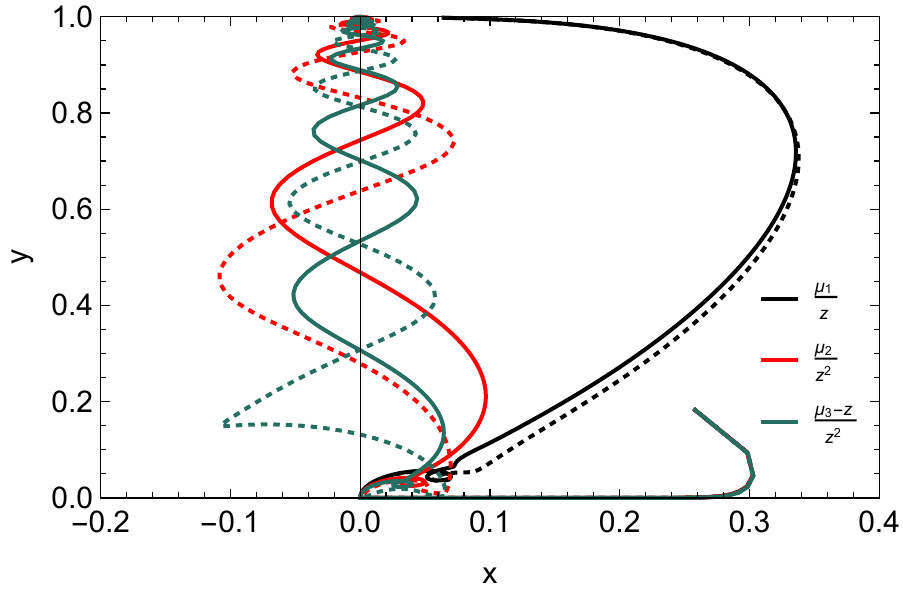}
\caption{{\bf Numerical peek at the cosmological dynamics II.} The plots show the evolution of the comoving Hubble radius with (top-right) and without (top-left) coupling, the evolution of the scalar field (bottom-left) and trajectories in the phase space (bottom-right) with (dashed) and without (regular) coupling.}
\label{plots1b}
\end{figure*}

We show the evolution of density parameters ($\Omega_r$, $\Omega_m$ and $\Omega_\phi$), density of the components, equation of state parameters $w_\phi$ and $w_{\rm eff}$ for the three cases $\mu_1/z^2$, $\mu_2/z$ and $(\mu_3-z)/z^2$ in Figure~\ref{plots1} with and without coupling. We also give the evolution of the comoving Hubble radius, the evolution of the scalar field and trajectories of the system in the phase space in Figure~\ref{plots1b}. The parameters are taken to be $\mu_1 = 1$, $\mu_2=39$ and $\mu_3 = 100$ and the coupling constant $\beta=0.1$ assuming a weak coupling. We evolve the system by setting the initial conditions at the epoch of nucleosynthesis which corresponds to $N\approx-22$ (although the plots for the $\Omega s$ are clipped to show from $N\approx-10$ since initially it is all radiation!) from temperature considerations. For the initial conditions, we choose $\Omega_{r_i} = 0.9$ and $\gamma_{\phi_i} = 4/3$ and an initial slope parameter $z_i$ so that the scalar field mimics radiation at the nucleosynthesis epoch. For determining initial normalized Hubble parameter $h_i=H_i/H_0$ and initial $\Omega_{\phi_i}$, we use $\Omega_{r_0}$ computed from $T_{\rm CMB} = 2.725$ K and take $ \Omega_{m_0} = 0.286$ \cite{omegam}. With this choice of initial conditions, we see that for all three cases, $f(z) = \mu_1/z^2$, $\mu_2/z$ and $(\mu_3-z)/z^2$, the dynamics of the scalar field is of a tracker type. This is particularly evident in the evolution of density in Figure~\ref{plots1} (middle-right) for $f(z) = (\mu_3-z)/z^2$. The density of the scalar field falls closely with radiation till it dominates and then scales with matter after radiation-matter equality and finally exits to give the late-time acceleration. The comoving Hubble radius (top rows in Figure~\ref{plots1b}) starts off as a line of unit slope featuring the radiation era which goes into the matter phase and then tilts down again with a unit slope implying accelerated expansion.  We also show the evolution of $w_\phi$ and $w_{\rm eff}$ for the three cases. The effect of coupling is seen to be more pronounced on $w_\phi$ than any other quantity. Such an effect was also shown in ref.~\cite{halo} in the case of specific potentials. It is interesting to note therefore if we can have any observables which are local in redshift, it can serve to constrain the model, $f(z)$ and the coupling. Unfortunately, the current observables like luminosity distances integrate out these features leading to degeneracy between many models.

Finally, we consider an example with $f(z) = \mu -z$ without coupling. For this case, noted earlier as well, we have $z_*= \mu$ and $z^2 f(z)|_{z=0} = 0$. We take different values for the parameter $\mu = 2$, $\sqrt{3}$, $\sqrt{2}$, $1$ and $-1$ to illustrate different aspects of evolution of the system as per the stability of the fixed points. We plot the phase space variables $(x,y,z,u)$ and $w_{\rm eft}$ for these cases in Figure~\ref{plots2}. The plots confirm the following aspects:
\begin{figure*}[t!]
\centering
\includegraphics[width=0.45\textwidth,scale=0.2]{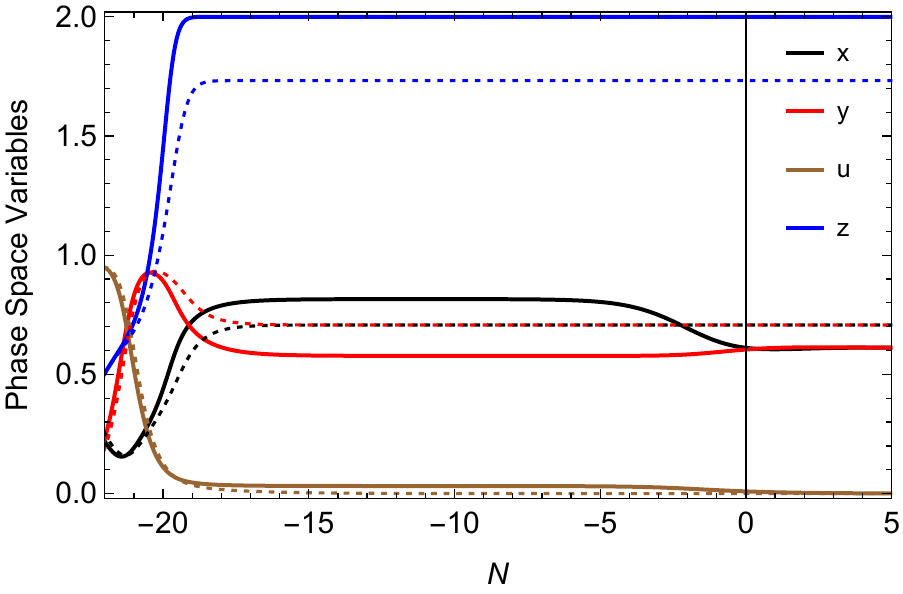}
\includegraphics[width=0.45\textwidth,scale=0.2]{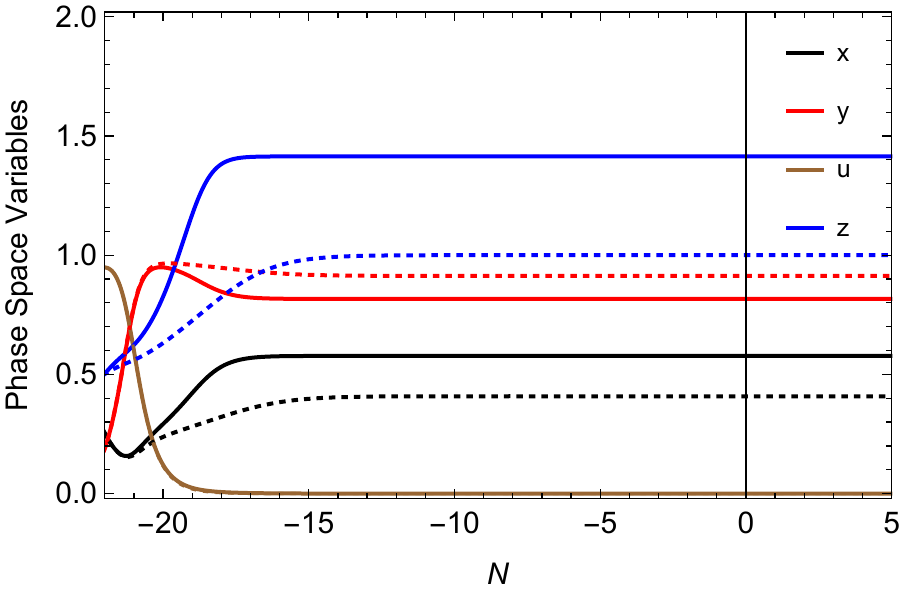}
\includegraphics[width=0.45\textwidth,scale=0.2]{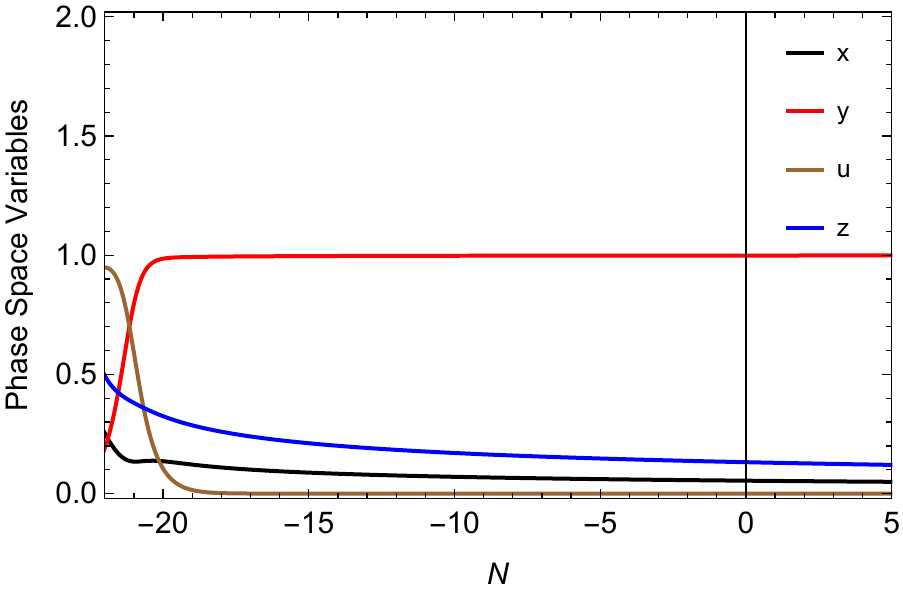}
\includegraphics[width=0.45\textwidth,scale=0.2]{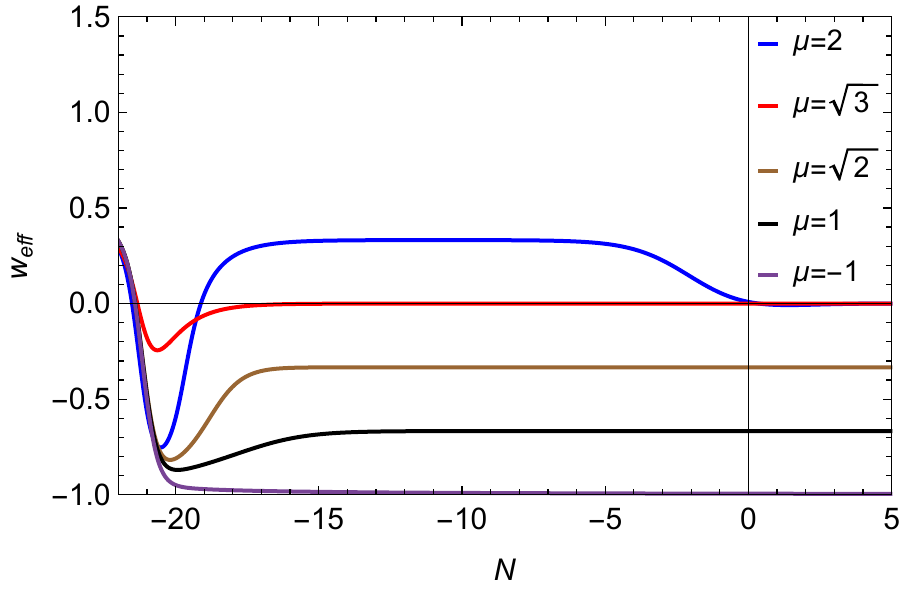}
\caption{The plots in this figure display the evolution of phase space variables and $w_{\rm eff}$ for $f(z) = \mu -z$ without coupling. We take different values for the parameter $\mu = 2$, $\sqrt{3}$, $\sqrt{2}$, $1$ and $-1$ to illustrate different aspects of evolution of the system as per the stability of the fixed points. The plots show and confirm the scaling behaviour, non-accelerated and accelerated phases and the de Sitter phase in the late-time evolution for these values of parameter $\mu$ which is the zero of $f(z)$.}
\label{plots2}
\end{figure*}

\begin{enumerate}
\item For $\mu = 2$, it is seen from the conditions of stability, that only point P$_{10}$ in Table~\ref{table1} is stable. Hence, we see in Figure~\ref{plots2} (Top-left), the phase space variables (regular lines) approach the values for P$_{10}$ with $x = y \approx 0.61$, $u \approx 0$ and $z = 2$ and $w_{\rm eff}$ (Bottom-right) goes to zero at late-times.

\item For $\mu = \sqrt{3}$, the system evolves to point P$_8$ in Table~\ref{table1} which is stable in this case and the scalar field mimics matter. The phase space variables (dashed lines) in Figure~\ref{plots2} (Top-left) approach the values for P$_{8}$ with $x = y \approx 0.707$, $u \approx 0$, $z = \sqrt{3}$ and $w_{\rm eff}\approx 0$ (Bottom-right) at late-times. There is no late-time acceleration in this case.

\item For $\mu = \sqrt{2}$, the stable point P$_8$ shows critical behaviour as $w_{\rm eff}\rightarrow -1/3$ at late-times. The phase space variables (regular lines in Figure~\ref{plots2}, Top-right) approach the values $x = 0.577$, $y = 0.816$, $u \approx 0$ and $z = \sqrt{2}$.

\item Point P$_8$ shows accelerated behaviour for $\mu = 1$. The phase space variables (dashed lines in Figure~\ref{plots2}, Top-right) approach the values $x = 0.408$, $y = 0.912$, $u \approx 0$ and $z = 1$ while $w_{\rm eff}\rightarrow -2/3$ at late-times.

\item Flipping the sign of $\mu$ with $\mu = -1$, point P$_{1}$ in Table~\ref{table1} becomes the stable node. Hence, we see in Figure~\ref{plots2} (Bottom-left), the phase space variables approach the values $x = 0$, $y = 1$, $u = 0$ and $z = 0$ asymptotically with $w_{\rm eff}\rightarrow -1$ giving asymptotic de Sitter behaviour. This can be tuned to give the correct cosmological parameters at the present epoch but in the `thawing' class of quintessence models with a shallow slope.
\end{enumerate}

The numerical computation of the dynamical equations thus validates the fixed point analysis to ascertain the background evolution of the quintessence models. The potential for the scalar field affects the background dynamics through the function, $f(z)$. It would be interesting to see if there is a way to reconstruct this function given the observational data.

\section{Summary and Discussion}
\label{conc}

`Quintessence' models explore the idea of using a scalar field as the dark energy component in the cosmological soup. The extra scalar degree of freedom also arises naturally if one considers modifications to general relativity in various ways and the resulting models can also pick up a coupling between the scalar field and matter fluid leading to a coupled dark sector dynamics. Between all this, the potential for the scalar field is motivated either phenomenologically or from higher dimensional theories. The question we ask ourselves is, ``Can we understand the background cosmological dynamics with a generic potential for the scalar field?''.

It turns out that it is indeed possible to do so in a way that can encompass a large class of potentials. The cosmological dynamical equations can be transformed into an equivalent autonomous system of differential equations. The non-linear system can be analysed analytically only through the linear stability analysis which includes computing the fixed points for the system and assess their stability. Even at this level, it is seen to be quite an effective tool for understanding background dynamics since these fixed points can be associated with various cosmological solutions that can decide the fate of the evolutionary dynamics.

We performed such an analysis by considering a \emph{three component system} with a scalar field, matter and radiation fluids but keeping the potential for the scalar field generic. We also considered a coupling between the scalar field and matter fluid of the form $Q = \sqrt{2/3}\beta\kappa\dot{\phi}\rho_m$ which can be motivated by considering a non-minimal coupling in the Jordan frame and then transforming the action to Einstein frame. It turns out that the equivalent dynamical system depends only on a function $f(z)$ of the relative slope of the potential which can be classified into three main categories for a large class of potentials beyond simple exponential type.  

We find that the fixed points for the three fluid system describe different cosmological solutions and can be associated with source/sink/transient phases of evolution depending on their conditions of stability. We see the existence of de Sitter dominated point, scaling solutions in which the scalar field behaves like radiation or matter thus leading to tracker type behaviour, matter/radiation dominated points. These points have different stability conditions and cosmological parameters. For some of the cases, the de Sitter point can become a global attractor and can be associated with the late-time acceleration. The effect of coupling renders few more scaling points and also modifications to the stability conditions of the persisting points. The coupling is introduced to account for the coincidence problem of $\Lambda$CDM model and allows for the interaction and transfer of energy between the dark matter and quintessence field. The analysis, to be stressed again, is done \emph{independent} of the form of the potential for the scalar field.

Finally, we used numerical techniques to illustrate and confirm the analysis by taking a few examples. We considered examples from each of the cases from the classification by taking the function $f(z)$ of the form $\mu/z^2$, $\mu/z$, $(\mu-z)/z^2$ and $(\mu-z)$ with and without coupling. For the choice of hand-picked initial conditions, we see the realistic evolution of the cosmological parameters for these cases. The first three examples show tracker type behaviour for the scalar field while the last one can be realised with the observed cosmological evolution only under thawing class of scalar field models. Further, within the last case, we illustrate the aspects of the fixed point analysis in different settings. We observe that the coupling shows major effect on the equation of state parameter of the scalar field and also allows for smoothening of the density evolution for the scalar field in the scaling regime. However, most of the current background-level cosmological probes integrate out these effects along the redshift leading to degeneracy between many models. It would be interesting to see how we can employ the parameter-estimation methods and study the perturbed evolution also in a generic way. This will have two-fold motive: (i) to study deviations from $\Lambda$CDM and (ii) if possible, reconstruct the potential or the function $f(z)$ for the quintessence field.  

\acknowledgments

Work of S.S. is supported by Dr. D. S. Kothari Postdoctoral Fellowship from University Grants Commission, Govt. of India. The authors also thank M. Sami and T. R. Seshadri for useful discussions and also acknowledge the services of IRC, Delhi where most of this work was done. Credits also to softwares Wolfram Mathematica and IPython notebook interface \cite{wpy} used for symbolic and numerical computations respectively. We also acknowledge the anonymous referee for useful comments.

\end{document}